# O-band QKD link over a multiple ONT loaded carrier-grade GPON for FTTH applications

N. Makris [1], A. Ntanos [2], A. Papageorgopoulos [1], A. Stathis [2], P. Konteli [1], I. Tsoni [1], G. Giannoulis [2], F. Setaki [3], T. Stathopoulos [3], G. Lyberopoulos [3], H. Avramopoulos [2], G. T. Kanellos [1,*], D. Syvridis [1]

[1] National and Kapodistrian University of Athens, Athens, Greece
[2] National Technical University of Athens, Athens, Greece
[3] COSMOTE Mobile Telecommunications S.A., Athens, Greece
*gtkanellos@di.uoa.gr

**Abstract:** We have successfully integrated an O-band commercial Quantum-Key-Distribution (QKD) system over a lit GPON testbed that replicates a carrier-grade Fiber-to-the-Home (FTTH) optical access network with multiple ONTs to emulate real-life FTTH operational deployments. © 2023 The Author(s)

## 1. Introduction

Quantum Key Distribution (QKD) has been proposed as a main solution against quantum threat, but for QKD technology to become economically viable in real life deployments, coexistence of quantum and classical channels should be facilitated by integrating QKD into existing infrastructure, reducing costs and complexity. However, to claim quantum-level security in an end-to-end network implementation, QKD should also be applied over the standardized fiber access networks, such as the Gigabit Passive Optical Network (GPON). Such implementations may be challenging due to the quantum channel sensitivity to transmission losses, impacting achievable distances and Secure Key Rates (SKR), as well as the significant losses incurred at the unavoidable splitting stages and the presence of excessive noise in the upstream/downstream channels.

Several studies have explored the integration of QKD into access networks [1]. In [2], Toshiba demonstrated a C-band QKD channel coexisting with a single O-band ONT over a 1:8 splitting stage. More recently in [3], a 10G-EPON system demonstrated the use of Discrete Variable QKD (DV-QKD) using a single feeder fiber and C-band quantum channel, while the upstream channels were in the O-band. The system demonstrated simultaneous operation of 3 Optical Network Units (ONUs) achieving an average SKR of 2.1 kbps per user over a 21 km transmission distance. In another study [4], QKD integration was showcased in GPON and NG-PON2 infrastructures using the Differential-Phase-Shift QKD (DPS-QKD) protocol. The GPON setup employed a dual feeder fiber with one fiber for the quantum channel (O-band) and the other fiber for the GPON upstream/downstream channels in the C-band. For the NG-PON2, the quantum channel was in the C-band, and the upstream channels were in the O-band. These configurations achieved an SKR of 0.51 kbps with a Quantum Bit Error Rate (QBER) as low as 3.28%.

Here, we present for the first time to our knowledge a complete coexistence scheme of an O-band QKD link over a single feeder fiber GPON implementation that is lit with up to 9 Optical Network Terminals (ONTs) to emulate a partial full load of a fully deployed and operational Fiber to The Home (FTTH) carrier-grade access network employing two splitting stages towards a 1:16 total splitting ratio, mirroring real-world deployments of the COSMOTE operator. We choose the quantum signal to co-propagate with the 2.5 Gbps 1490nm downstream GPON channel and operate in the O-band to suppress the Raman scattering effects, while counterpropagating in the fiber with the O-band GPON upstream channels with a spacing of just 4-6 nm from the quantum channel. We observe a ~3 dB SKR degradation with one ONT in operation mainly due to the contribution of the back-reflected photons from the ONT leaking from the bandpass filter (BPF) of the QKD receiver (Bob). We further evaluated the lit GPON operation allowing for up to 9 ONTs in the wavelength range from 1314 nm to 1316 nm to operate for approximately 60 hours, allowing user rates up to 500 Mbps in each ONT. Surprisingly, we noted an improvement in the performance of the QKD link in terms of SKR when adding more ONT users due to the Power Levelling Sequence upstream (PLSu) function of GPON [5] to reduce the launching power of the ONT.

## 2. Experimental Setup

A sustainable QKD link is demonstrated between the quantum transmitter (Alice) residing on the NOKIA 7360 ISAM GPON Optical Line Terminal (OLT) [5] side (central office) of the GPON and the quantum receiver (Bob) residing on the ONT side (end-user). The GPON network testbed, as illustrated in Fig. 1, emulates a real-life FTTH setup of COSMOTE telecom operator using commercial off-the-shelf components and 4 km fibers with a total link attenuation of 21 dB. For the QKD pair we used a Toshiba QKD4.2-MU/MB pair [7] that implements the Toshiba T12 protocol (efficient BB84 protocol with decoy states and phase encoding) [6] with a link budget of 24 dB. The 1310 nm QKD quantum channel co-propagates with the Class C+ 10 Gigabit Small Form-factor Pluggable (XFP) transceiver 1490 nm downstream GPON channel from the OLT with +3 dBm transmission power and 30dB link budget and the 1529nm and 1530nm service channels from Toshiba. Following the standard 2-stage splitting stage deployment of COSMOTE operator in their commercial GPON, these signals propagate through a 3 km G652D fiber spool that covers the physical distance between the Central Office (CO) and the Cabinet (CAB). The signals are then split in the first 1:4 splitting stage and connect to the CAB and the Building Entry Point (BEP) through a 1 km G657A1. At BEP the signals are split again in the second 1:4 splitting stage and after propagating for another 40 meters are finally split through a 1:2 splitter to ensure that both signals are efficiently coupled and both reach the ONT and the quantum receiver (Bob). For the upstream configuration, the ONTs transmit signals with wavelengths that vary from 1314 nm to 1316 nm.

In our setup, the back reflections of the ONT signal in the second 1:4 splitting stage and the 1:2 coupler before Bob are the main limiting factors to obtain a successful QKD link. To compensate for this, we tune the nominal 0 dBm transmission power of the Class B+ XFP Nokia G-010G-R ONTs upstream channels down to -3 dBm with no penalty in the ONT link performance. Moreover, as we could not initiate a QKD link when the ONT transmission wavelength was below 1312nm and observed that the optimum SKR results are acquired when we ONTs operate at the furthest possible spacing (1316 nm) from the QKD channel (1310nm), we employed only ONTs emitting in the 1314-1316 nm. The rest of the ONTs are connected to the remaining ports of the 1:4 BEP splitter and the last 5 ONTs through an additional 1:8 second stage splitter. Toshiba requires the third service channel with a direction from Bob to Alice, that was implemented as a back-to-back (10 m) connection since our laboratory experiments have demonstrated minimal influence of the fiber link. After effectively managing the emission powers of the GPON signals we progressively deployed the 9 ONTs ensuring every time that the link is operational and Secure Key Generation occurs. In three ONTs we attached IP routers to evaluate the high-speed internet connection.

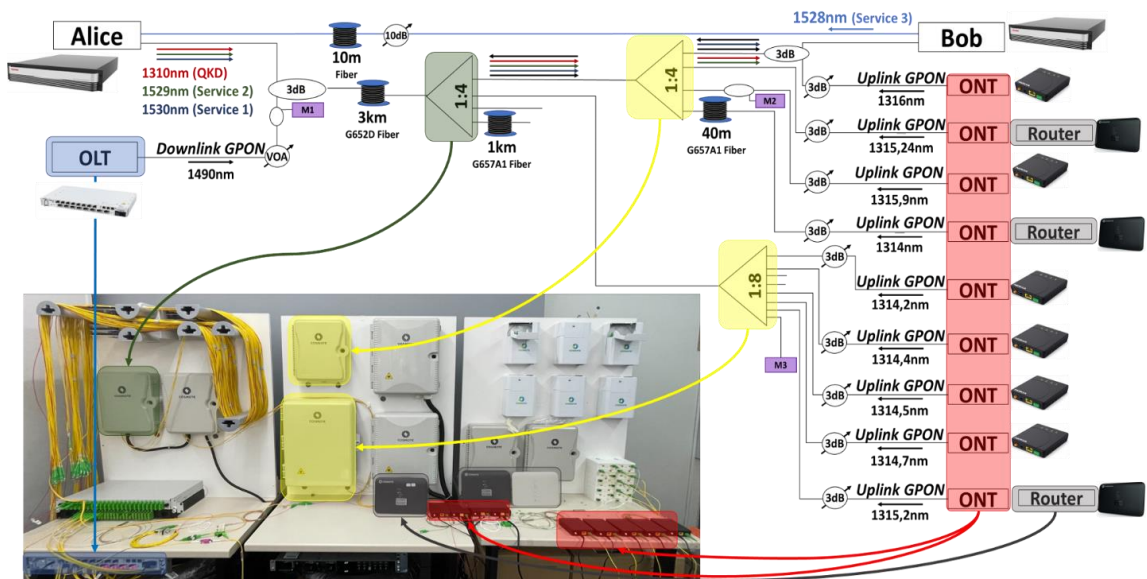

Fig. 1: The GPON topology, along with the QKD setup, presented both as a diagram and in its real counterpart.

3. **Results and Discussion**

We initially apply the QKD link over the dark GPON setup, hence the SKR performance is affected only by the splitting stages and the 21 dB link loss, leading to a 3.13% QBER and 20 kbps SKR, as shown in Table 1. After initiating the first ONT (1316 nm) operation, the performance of the QKD over the lit GPON is degraded by more than 3 dB in the SKR that is attributed to the in-band noise from the back reflections of the splitting stages, as well as the Raman scattering of the 1490nm downstream. Similar SKR and QBER degradation occurs after the deployment of 3 additional ONTs on the ports of the 1:4 BEP splitter. However, remarkably enough, there is an improvement of the QBER and SKR after the deployment of the 5 additional ONTs (9 in total) on the 1:8 BEP splitter. This is due to the PLSu function of GPON that assists in the adjustment of ONU power levels to reduce the optical dynamic range as seen by the OLT and the fact that the multiplexing method employed is Time Division Multiplexing (TDM) with Dynamic Bandwidth Allocation (DBA). When the OLT handles a reduced load of ONTs, higher powers are employed for communication between the OLT and the ONT. Upon deploying the additional 5 ONTs, we noted a 3.2 dB drop in total signal power which corresponds to 0.6 dB per ONT. Consequently, a greater number of back reflected photons are induced in Bob QKD receiver when less ONTs are employed, as shown in Table 1 with the back reflections counting -64.3 dBm. In the case of the 9 ONTs, an improved back reflection level at –67.1 dBm led to an improved QBER and SKR performance respectively. These ONTs were attached to IP routers to confirm that they delivered a consistent high-speed internet connection, achieving speeds up to 500 Mbps. Finally, the QKD over GPON coexistence testbed was able to run uninterrupted for a time period of 60 hours, with a total number of 9 ONTs loaded as shown below in Fig. 2.

Table 1: Average QBER and SKR results for the 4 different load setups.

| # of ONTs | QBER | SKR (bps) | Back reflections |
|---|---|---|---|
| No ONTs | 3.33 % | 20 K | - |
| 1 ONT | 5.8% | 9 K | -64.3 dBm |
| 5 ONTs | 6.15% | 6 K | -65.4 dBm |
| 9 ONTs | 5.11% | 10.1 K | -67.1 dBm |

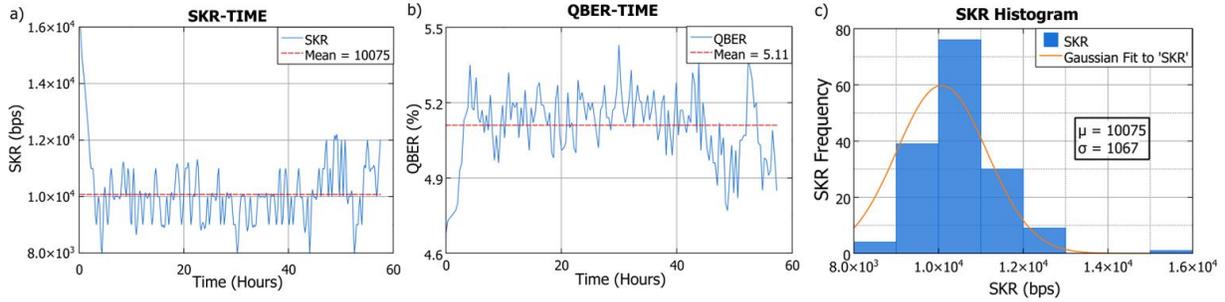

Fig. 2: Metrics Analytics a) SKR vs Time b) QBER vs Time c) SKR Histogram.

## 4. Conclusion

We have successfully demonstrated an O-band QKD link over a single feeder lit GPON network that simultaneously serves 9 users. A noteworthy observation was that deployment of additional ONTs positively impacted the QKD performance due to the power management of GPON systems. Operating for more than 60 hours, we achieved an average SKR of 10.07 kbps, and a QBER of 5.11%.

## 5. Acknowledgements

This work was funded by the EU quantum flagship project QSNP (GA 101114043) and the HellasQCI project (GA 101091504).